\documentclass[fleqn,usenatbib]{mnras}

\usepackage{newtxtext,newtxmath}

\usepackage[T1]{fontenc}

\DeclareRobustCommand{\VAN}[3]{#2}
\let\VANthebibliography\thebibliography
\def\thebibliography{\DeclareRobustCommand{\VAN}[3]{##3}\VANthebibliography}

\usepackage{graphicx}	
\usepackage{amsmath}	
\usepackage{subfig}

\title[Pulsars in the death valley]{Emission properties of 5 pulsars in the death-valley and implications on death line models}

\author[Singh et al.]{
S.~Singh,$^{1}$\thanks{E-mail: shubham.singh@manchester.ac.uk}
J.~Roy,$^{2}$
B. Bhattacharyya$^2$
\\
$^{1}$Jodrell Bank Centre for Astrophysics, University of Manchester, Manchester M13 9PL, Uk \\
$^{2}$National Centre for Radio Astrophysics, Tata
Institute of Fundamental Research, Pune 411 007, India\\
}

\date{Accepted XXX. Received YYY; in original form ZZZ}

\pubyear{2015}

\begin{document}
\label{firstpage}
\pagerange{\pageref{firstpage}--\pageref{lastpage}}
\maketitle

\begin{abstract}
In the framework of the coherent curvature radiation model of pulsar radio emission, charged particles responsible for the radio emission are generated on the polar cap in the localized pair cascade processes called sparks. When a pulsar can no longer sustain the sparking process on its polar cap, the coherent radio emission from the pulsar stops, and the pulsar is called dead. In this work, we revisit the pulsar death phenomena under two popular voltage gap models: vacuum voltage gap and partially screened voltage gap.  We notice that a dying pulsar resorts to a single spark on the polar cap to sustain the pair production under both voltage gap frameworks. The presence of only one spark on the polar cap has important implications for the radio emission properties of the pulsar. We study the emission properties of five pulsars close to the lower boundary on the $\mathrm{P}-\dot{\mathrm{P}}$ plane of the current pulsar population, as these pulsars are expected to be dying. We find that the dying pulsars in our sample and the normal pulsar population have very similar emission properties. We show that the majority of pulsars in the pulsar death valley, including five pulsars in our sample, show evidence of multiple sparks as opposed to what is expected from single spark death line models.

\end{abstract}




\section{Introduction}
The canonical pulsars slow down with time, and the spin-down rate ($\dot{\mathrm{P}}$) also decreases; as a result, pulsars move towards the lower right corner of the $\mathrm{P}-\dot{\mathrm{P}}$ plot \citep{ppdot_evolution_Johnston}. Pulsars with long periods and low period derivatives ($\dot{\mathrm{P}}$) seem to be missing from the current pulsar population in the bottom right corner of the $\mathrm{P}-\dot{\mathrm{P}}$ plane (see Fig. \ref{fig:death_valley}). Several observational and technical biases can contribute to the reduced density of observed pulsars in this region of the $\mathrm{P}-\dot{\mathrm{P}}$ plane. As per the conventional picture, the opening angle of the pulsar radio beam gets smaller with the increasing period, reducing the chances of it being visible from the Earth \citep{Maciesiak_2012, LBL_interpulse, beam_evolution_young}. It is also proposed that the magnetic axis aligns with the rotation axis of the pulsar with increasing age \citep{ppdot_evolution_Johnston, beam_evolution_young}. This alignment can significantly reduce the chances of our line of sight passing through the pulsar's radio beam. Along with these disadvantages caused by the pulsar's geometry, the pulsar radio luminosity is also expected to decrease with its increasing age \citep{luminosity_lyne}. Additionally, the conventional periodicity search methods have been shown to be heavily biased against the long period and small duty cycle pulsars \citep{GHRSSIII, ffa_morello, search_framework}. All these factors make it very difficult for pulsar surveys to detect the older population of long-period pulsars. In addition to the observational difficulties mentioned above, coherent curvature radiation, one of the popular models for pulsar radio emission, also predicts an end to radio emission from pulsars located in the lower right corner of the $\mathrm{P}-\dot{\mathrm{P}}$ plane \citep{single_park, chen_and_rutherman}.

The charged particles responsible for the pulsar radio emission are generated on the polar cap of the neutron star in the localized pair cascade processes (\citealt{RS75}, RS75 hereafter, and \citealt{Gil_2000}, GS00 hereafter). This pair cascade process, also known as spark, requires favorable physical conditions on the polar cap. These conditions include a sufficient voltage gap and curved magnetic field lines on the polar cap (e.g. \citealt{single_park}, \citealt{chen_and_rutherman}, and RS75). Pulsars that can not provide a sufficient voltage gap on the polar cap (due to slow rotation and/or weak magnetic field) can not sustain the sparking process and hence can not produce radio emission. This process of permanent cessation of radio emission is called the death of the radio pulsar. The observed lower boundary on the $\mathrm{P}-\dot{\mathrm{P}}$ diagram of the current pulsar population is thought to be a reflection of this limiting condition (\citealt{single_park}, \citealt{chen_and_rutherman}, and \citealt{young_1999}). Several attempts have been made to model the physical conditions responsible for pulsar death and predict the theoretical lower boundary or death line for the observed pulsar population (\citealt{deathline_beskin2}, \citealt{deathline_beskin1}, \citealt{single_park}, \citealt{chen_and_rutherman}, and RS75). These models depend on parameters that can vary from pulsar to pulsar (e.g. pulsar geometry and magnetic field configuration near stellar surface), hence instead of a rigid death line on the  $\mathrm{P}-\dot{\mathrm{P}}$, a region on the  $\mathrm{P}-\dot{\mathrm{P}}$ plane is more useful. This region, covering possible pulsar to pulsar variations in the model parameters, is called death valley \citep{deathline_beskin2, chen_and_rutherman}.

The death line models are mostly concerned with the voltage gap on the polar cap and suitable conditions for the pair cascade in it. RS75 proposed a vacuum gap model where a vacuum gap forms above the polar cap corresponding to the open field lines. This vacuum gap grows until the pair cascade starts and screens the voltage gap. The pair cascade happens in localized sparks that have dimensions equivalent to the gap height (GS00). This model was very successful in explaining various emission features of pulsars, including subpulse drifting. However, the subpulse drifting rates predicted by the vacuum gap model were higher than the observed rates. \citet{Gil_2003} proposed a partially screened gap model, that considers a steady outflow of ions from the polar cap, partially screening the voltage gap. This partial screening of the voltage gap provided a natural means to account for the slow subpulse drifting rates observed in pulsars. \citet{single_park} studied the phenomena of pulsar death within the framework of a partially screened gap (PSG). They concluded that a pulsar stops its radio emission once it is not able to sustain even a single spark on its polar cap. Hence, a pulsar that is about to die has only a single spark. \citet{single_park} support this single spark death line model using the emission properties of 8.5 s pulsar J2144-3933. On a closer look, the vacuum gap model also expects a single spark in dying pulsars due to insufficient voltage difference in its voltage gap. A vacuum voltage gap will keep growing until the gap height becomes equivalent to the radius of the polar cap. In this situation, the maximum possible voltage difference is available in the voltage gap \citep{chen_and_rutherman}. The spark formed in this case will be as large as the polar cap itself, hence only a single spark can be accommodated on the polar cap.

The pulsar radio beam and its evolution with the pulsar's age are not completely understood. Apart from the beam getting narrower and the magnetic axis aligning with the rotation axis, the complexity of the pulsar profile seems to evolve with age. It has been observed that young pulsars tend to have simpler profiles while the older pulsar population has more complex profiles \citep{profile_complexity_ArisK}. \citet{Profilecomplexity_vivekananda} suggested that either surface irregularity on the polar cap or the multipole component of the magnetic field is responsible for the complexity of profiles in the older population. While, \citet{profile_complexity_ArisK} proposed that the older pulsars emit from a wider range of heights in comparison to the younger counterparts, increasing the complexity of the observed profiles. However, analysis of a larger sample is required to robustly establish this empirical relation between profile complexity and pulsar age. The distribution of sparks on the polar cap largely decides the radio beam of the pulsar (RS75, GS00). The radio beam combined with the geometry of the pulsar decides the shape of the pulsar profile. The presence of only a single spark on the polar cap means that the radio beam will have only a single component. The folded profile in such a case will have only a single component irrespective of the geometry of the pulsar (though it can have an interpulse for orthogonal geometry). \citet{single_park} argued that a single spark will result in a single component profile and such pulsars can not show emission features like subpulse drifting and mode changing.

In this work, we revisit the single spark model of pulsar death in both vacuum gap and partially screened gap framework. We study a sample of five long-period pulsars close to the lower boundary on $\mathrm{P}-\dot{\mathrm{P}}$ diagram of the current population. We study the emission properties of these pulsars to find whether pulsars close to the death line are different from the normal pulsars, in terms of emission properties. We use the five pulsars in our sample, along with several other pulsars in the death valley to test the consistency of the single spark death models. We revisit the voltage gap models along with the single spark model of pulsar death in section \ref{sec:theory}. We describe our sample of death line pulsars in section \ref{sec:sample}. Section \ref{sec:similarity} describes the similarity between the normal pulsar population and the death line pulsars, in terms of emission properties. Then, in section \ref{sec:profiles}, we infer the minimum number of sparks on the polar cap of many death line pulsars and compare it with the number of components in their profiles. Finally, the outcome of this work and its implications on the pulsar death models are discussed in section \ref{sec:discussion}.

\section{Single spark model of pulsar death}\label{sec:theory}
In this section, we provide a brief overview of the sparking process, the two voltage gap models, and the pulsar death phenomena in these voltage gap models. First, we summarize the sparking process on the polar cap as per RS75. Then, following GS00, we describe the vacuum gap model and how dying pulsars are expected to have a single spark on their polar cap in this framework. Then we briefly describe the partially screened voltage gap model and single spark model of pulsar death following \citet{single_park}.  

\subsection{Voltage gap and sparking process on the polar cap}
According to RS75, neutron stars with an obtuse angle between their rotation and magnetic axes are expected to have positively charged magnetospheric plasma above their polar caps. This plasma can escape from the system through open field lines. As the neutron star can not provide ions (due to high binding energy) to replace the escaped charge, the positively charged magnetosphere above the polar cap separates from the neutron star surface and a voltage gap is formed. The voltage gap keeps growing to a point when the voltage difference across the gap is sufficient for a pair cascade to happen. This pair cascade produces electron and positron pairs that end up screening the voltage gap. Once the voltage gap is screened, the pair cascade stops and access charged particles escape from the system via open field lines. Hence, the voltage gap grows again and the same cycle repeats.\\
The separation between the polar cap surface and the positively charged plasma moving away from the surface is called the gap height and is denoted by $\mathrm{h}$. The following equation gives the maximum potential difference available across the gap (Eq. 15 of RS75),

\begin{equation}
    \Delta V = \frac{\Omega B_s}{c} h^2
\end{equation}

Here $\Omega$ is the angular velocity and $B_s$ is the magnetic field strength near the surface of the polar cap. For a pair cascade to happen within the gap, a photon produced by the curvature radiation of the primary particles should be able to produce secondary pairs within the gap height. The potential difference on the polar cap is responsible for accelerating the primary particles. These particles will attain a Lorentz factor given by the following expression,

\begin{equation}
\gamma = \frac{\mathrm{e}\Delta \mathrm{V}}{\mathrm{m}\mathrm{c}^2}  
\end{equation}

The particles with Lorentz factor $\gamma$ will produce high-energy photons via curvature radiation. The energy of these photons can be expressed as following (Eq.18 of RS75),
\begin{equation}
\hbar \omega \thickapprox \frac{3}{2}\gamma^3\hbar\mathrm{c}/r_c 
\end{equation}  
Here, $r_c$ is the radius of curvature of the magnetic field lines and $\gamma$ is the Lorentz factor of the charged particles (electrons/positrons). The condition for this photon in a strong magnetic field to produce the secondary electron-positron pair is the following (Eq. 3 of \citealt{chen_and_rutherman}),
\begin{equation}
    \frac{\hbar \omega}{2\mathrm{m}\mathrm{c}^2} \frac{B_{\perp}}{B_q} \thickapprox \frac{1}{15} 
\end{equation}

Where, $$B_q = \frac{\mathrm{m}^2 \mathrm{c}^3}{\mathrm{e}\hbar} = 4.4 \times 10^{13}~\mathrm{Gauss}$$ and  
$\hbar \omega$ is the photon's energy, $\mathrm{m}$ is the mass of an electron, and $B_{\perp}$ is the magnetic field strength perpendicular to the motion of the photon. These photons are generated by the curvature radiation from relativistic positrons/electrons. 
The maximum achievable angle between the direction of the photon's motion and the magnetic field within the voltage gap is $h/r_c$ and hence the maximum value of $B_{\perp}$ will be,
$$B_{\perp} \thickapprox B_s h/r_c$$

The condition for pair cascade can be obtained by combining the equations 2, 3, and 4. This results in an expression similar to the equation (21) of RS75,
\begin{equation}
    \left( \frac{3\hbar}{4\mathrm{m}\mathrm{c}r_c} \right) \left( \frac{\mathrm{e}\Delta V}{\mathrm{m}c^2}\right)^3 \left( \frac{B_s \mathrm{h}}{r_c B_{\perp}}\right) \thickapprox \frac{1}{15}
\end{equation}
 If the voltage gap on the polar cap gets wide and strong enough to fulfill this condition, several localized pair cascades will take place, and the gap will be screened by the charged particle produced in the cascade. These charged particles will escape the gap in a timescale of $h/c$ and the potential gap will start growing again. This repetitive process of localized pair cascade is called spark on the polar cap. A spark happening at a particular location screens the voltage gap around it and prevents another spark from occurring within a distance equivalent to the photon mean free path (RS75). GS00 argue that sparks are two-dimensional entities on the polar cap with a characteristic dimension similar to the gap height $\mathrm{h}$. These sparks are separated from each other by a distance $\mathrm{h}$ (see Fig. 1 of GS00).

\subsection{Pulsar death-lines in vacuum gap models}
The height of the voltage gap ($\mathrm{h}$) can grow only up to the value of the polar cap radius ($\mathrm{r_p}$). If the pulsar parameters are such that a condition for pair cascade can not be satisfied even at voltage gap height approaching the value of the polar cap radius, then the plasma column above the polar cap flows away (RS75). The maximum voltage difference across the gap in such a case is given by,

\begin{equation}
    \delta V_{max} = \frac{\Omega B_s}{c} {r_p ^2} = \frac{\Omega \Phi}{c}
\end{equation} 

Here, $\Phi$ is the flux of the open magnetic field lines. This flux is only sensitive to the dipolar component of the surface field strength (RS75). However, the multi-polar surface field still decides the radius of curvature of field lines and the strength of the magnetic field at the location of the spark. \\
\cite{chen_and_rutherman} derived a set of death lines on the $\mathrm{P}-\dot{\mathrm{P}}$ plane using a voltage difference of $\Delta V_{max}$ in the vacuum gap, and different combinations of surface magnetic field strengths and curvature radius for the field lines in equation (5). A pulsar can die anywhere in between the death line for pure dipolar field configuration (see equation (6) of \citealt{chen_and_rutherman}) and the death line for extremely multipolar and curved field lines at the polar cap (see equation (9) of \citealt{chen_and_rutherman}). The part of the $\mathrm{P}-\dot{\mathrm{P}}$ plane in between these two death lines is known as death-valley. It should be noted that the maximum voltage difference ($\Delta V_{max}$) is achieved only when gap height ($h$) approaches the value of polar cap radius ($r_p$).  In this configuration, the voltage gap can be screened with only a single spark because the photon mean free path is equivalent to the polar cap radius. Hence, in this death line model, the polar cap will have only a single spark before completely stopping its radio emission. \\

A set of single spark death lines can be derived in the context of the vacuum gap model using the complexity parameter ($a$) defined in GS00. This complexity parameter is the ratio of the polar cap radius ($r_p$) and the diameter of a spark ($\mathrm{h}$), given by the following equation,
\begin{equation}
    a_{vg} = \frac{r_p}{h} \thickapprox 5 \times 10^4 \left( {R_6}^{-2/7} b^{1/14}\right) \times \dot{\mathrm{P}}^{2/7} \mathrm{P}^{-9/14}
\end{equation}

Here, $R_6 = r_c/10^{6}$ cm, represents the radius of curvature of the magnetic field lines at the polar cap surface, and $\mathrm{b}$ is the ratio of multipolar field strength at the surface and the dipolar field strength inferred from observations. It should be noted that $\mathrm{r}_\mathrm{c} \sim 10^{6}$ cm implicitly assumes that the magnetic field curvature is highly multipolar near the stellar surface \citep{RS75}. GS00 argue that the factor $ \left( {R_6}^{-2/7} b^{1/14}\right) \sim 2$ for all realistic cases of the multipolar field configuration. If we use $\dot{\mathrm{P}} = \dot{\mathrm{P}}_{-15} \times 10^{-15} ~s/s$ then expression for complexity parameter becomes,
\begin{equation}\label{eqn:avg}
\begin{split}
    a_{vg} = \frac{r_p}{h} \thickapprox 5 \times \left( 0.517 \times {R_6}^{-2/7} b^{1/14}\right) \times \dot{\mathrm{P}}_{-15}^{2/7} \mathrm{P}^{-9/14} \\
    = 5 \times C \times \dot{\mathrm{P}}_{-15}^{2/7} \mathrm{P}^{-9/14}  
\end{split}
\end{equation}
Where, $C =  \left( 0.517 \times {R_6}^{-2/7} b^{1/14}\right)$ is a parameter that solely depends on multipolar field structure on the polar cap and should be $\sim 1$ for all realistic cases of multipolar field structure. \\
For a pulsar to emit in radio, the complexity parameter ($a_{vg}$) must allow at least one spark on the polar cap. So, a death line can be obtained by equating the complexity parameter to 1 (i.e. the polar cap will have a single spark),
\begin{equation}
    \dot{\mathrm{P}}_{-15} = \left(  {5 \times 0.517 \times R_{6}^{-2/7} b^{1/14}}\right)^{-7/2} \mathrm{P}^{9/4}
\end{equation}
This death line is a single spark death line that doesn't only consider the feasibility of the pair cascade process but also the sustainability of at least a single spark on the polar cap. Notably this single spark death line under the framework of the vacuum gap model has a weak dependence on the actual field configuration at the polar cap. The field configuration at the polar cap is characterized by the ratio of the multipolar field and dipolar field strengths ($\mathrm{b}$) and curvature of the field lines at the polar cap ($R_6$). We plot the death valley for this death-line model, the region on $\mathrm{P}-\dot{\mathrm{P}}$ diagram between death lines for two cases of magnetic field configuration, moderately mulitpolar field configuration, and a realistic case of highly multipolar field configuration (blue shaded region in Figure \ref{fig:death_valley}). We use $R_6 = 1.0$, $\mathrm{b}=1$ for moderately multipolar field configuration (similar to case (2) of \citealt{chen_and_rutherman}) and $R_6 = 0.1$, $\mathrm{b}=100$ for extremely multipolar field configuration. Due to the weak dependence of the complexity parameter ($a_{vg}$) on the surface field configuration (GS00), the death lines for this model do not change much between the two cases of magnetic field configuration. Hence, the resulting Death Valley is narrow.

\subsection{Partially screened gap (PSG) model}
The partially screened voltage gap (PSG) model is a variant of the vacuum voltage gap proposed by \citet{Gil_2003}. This model considers a steady stream of outflowing ions from the polar cap. As a result, the voltage gap above the polar cap is no longer a pure vacuum gap but is partially screened by the steady stream of ions. Here, we briefly summarize the PSG following \citet{single_park}.\\

When a spark happens on the polar cap, the electrons produced in the pair cascade travel towards the stellar surface and hit the surface resulting in the heating up and increase in the surface temperature ($T_s$). Once this temperature $T_s$ approaches a temperature equivalent to the binding energy of ions ($T_i$), positive ions from the polar cap surface start flowing, partially screening the voltage gap. The expression for ion charge density ($\rho_i$) in the voltage gap is given by the following expression \citep{single_park},

\begin{equation}
    \frac{\rho_i}{\rho_{GJ}} = \mathrm{exp} \left( 30 - \frac{T_i}{T_s}\right)
\end{equation}
Where $\rho_{GJ}$ is the Goldreich and Julian charge density, required to screen the gap completely (RS75). This flow of ions (with ${\rho_i} < {\rho_{GJ}}$) partially screens the voltage gap,

\begin{equation}
    \Delta V_{PSG} = \eta \Delta V_{VG}
\end{equation}
Here, $\eta = 1 - \rho_i/\rho_{GJ}$ is the screening factor. This partially screened gap model successfully accounts for the observed X-ray emission from pulsars and the timescales of the observed subpulse drifting \citep{Gil_2006}.
\subsection{Death-line in PSG models}
\begin{figure*}
    \centering
    \includegraphics[width=0.6\linewidth]{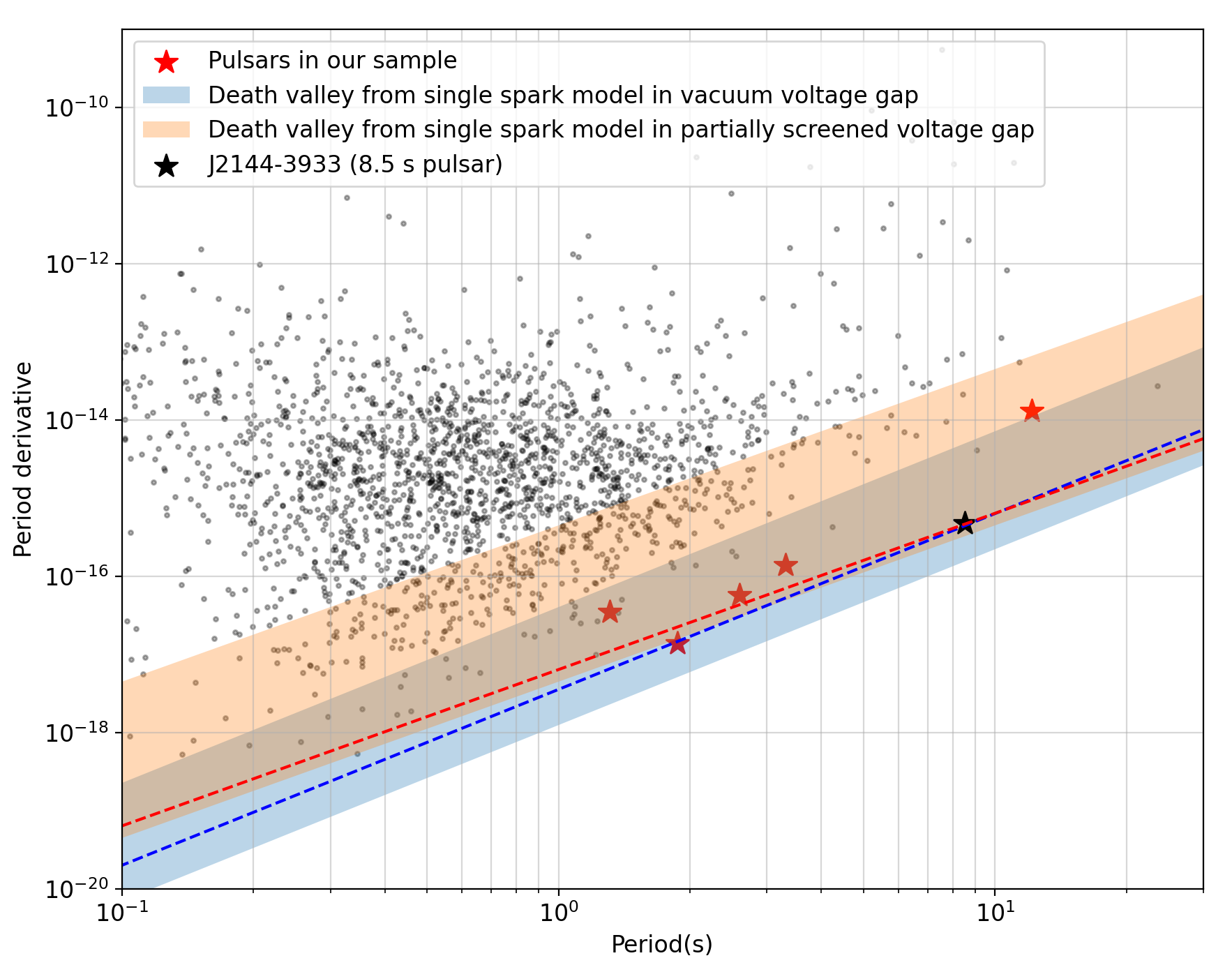}
    \caption{Selected pulsars on the $\mathrm{P}-\dot{P}$ plane as red stars. Shadowed regions represent the death valley of two single spark models. The blue-shaded region shows the single spark death valley in the case of the vacuum gap model, while the yellow-shaded region shows the death valley in the case of the single spark model in the PSG framework. The red dashed line is the single spark death line under the PSG framework passing through 8.5 s pulsar (J2144-3933, black star), while the blue dashed line is the single spark death line for the same pulsar under the vacuum gap framework.}
    \label{fig:death_valley}
\end{figure*}

\citet{single_park} derived an expression for the ratio of spark area and polar cap area under the framework of PSG. The inverse of this ratio can be considered as the 2D equivalent of the complexity parameter discussed in earlier. 

\begin{equation}\label{eqn:apsg}
    a_{psg} = \frac{A_{pc}}{A_{sp}} = \frac{1}{3.16 \times 10^{-4}} \frac{\eta^2 b \cos^2\alpha_l}{T_6^4}\frac{\dot{\mathrm{P}}_{-15}}{\mathrm{P}^2}
\end{equation}
Here, $\eta$ is the screening parameter defined in equation (11), b is the magnetic field configuration parameter defined earlier, $\alpha_l$ is the angle between the local surface magnetic field and the rotation axis, and $T_6$ is the surface temperature ($T_s$) in the units of $10^6~K$. \\
A single spark death line in this case can be obtained simply by equating $a_{psg}$ to 1 (see equation (5) of \citealt{single_park}). One of the main differences between parameter $a_{vg}$ and parameter $a_{psg}$ is their dependence on the parameter b ($B_s/B_p$). The parameter $a_{psg}$ strongly depends on b; hence we can expect a wide death valley under the PSG framework. The orange-shaded region in Figure \ref{fig:death_valley} shows the death valley for a single spark model in the PSG framework. We use $\eta = 0.15$,
 $T_6=2$, and $\alpha_l \sim 45^{\circ}$ as used by \citet{single_park}. We use $\mathrm{b}=1$ for dipolar field strength (upper limit on the death-valley) and $\mathrm{b}=100$ for the highly multipolar field configuration (lower limit on the death-valley).

\section{Sample selection, observations, and detections}\label{sec:sample}

We aim to study pulsars near the observed lower boundary on the $\mathrm{P}-\dot{\mathrm{P}}$ plane. As discussed earlier, this lower boundary can be linked to the expected smaller inclination angle, lower radio luminosity, and narrower radio beam of the older pulsar population, limiting our chances of discovering this population. The limited sensitivity of pulsar surveys and possibly biased search methods can add to the low number of discoveries in this parameter space. However, the coherent curvature radiation model for pulsar radio emission interprets this observed lower boundary on the  $\mathrm{P}-\dot{\mathrm{P}}$ plane as a result of the pulsar death phenomena. We selected five pulsars that are close to the lower boundary of the current pulsar population on the $\mathrm{P}-\dot{\mathrm{P}}$ plane (see Fig. \ref{fig:death_valley}). The table \ref{tab:source_list} lists the pulsars in our sample and their parameters. All the pulsars in our sample are long-period and low-DM pulsars. We observed these pulsars with band-4 (550 - 750 MHz) of the uGMRT \citep{GMRT_GWB}. We used the newly implemented real-time RFI filtering \citep{realtime_RFI} during these observations. All the observations were done in phased array (PA) mode, but we also recorded the incoherent array (IA) mode data in the background for a few observations. We use these IA beam observations to perform beam subtraction (PA - IA, \citealt{post_corr}) which removes the dominant RFI from the beam data. \\
We detected all pulsars in our sample. Since these are long period and low DM pulsars, we could not use the zero DM subtraction to eliminate the broadband RFI. We performed beam subtraction for observations when the IA beam was recorded in the background, and we found that the beam subtraction method effectively removes almost all broadband and narrowband RFI features from the data. The broadband RFI features could not be removed in some observations when we did not record IA beam data in the background. Performing the beam subtraction seems necessary when dealing with long period and/or low DM pulsars.\\
Two of our detections had dominant baseline variations: the observation of pulsar J0343-3000 on 7 December 2021 and the observation of pulsar J1503+2111 on 14 November 2023. To remove the baseline, we use the following procedure: We cut out the on-pulse window and replace it with zeros, then use running median subtraction to remove baseline variations, and then add the on-pulse window back. We use a running median window of size comparable to the on pulse duration of these pulsars.

\begin{table*}
    \centering
    \begin{tabular}{lccccccr}
      Pulsar name & DM & Period & Period-derivative & $a_{vg}$ & $a_{psg}$ \\
                & $\mathrm{pc}-\mathrm{cm}^{-3}$ & (s) & ($\times 10^{-15}~s/s$) & ($r_p/h$, vacuum gap) & ($A_{pc}/A_{sp}$, PSG)\\
                \hline
     J0343-3000 & 20.2 & 2.597 & 0.04 & 1.03 &  0.84\\
     J1115-0956 & 16.1 & 1.310 & 0.034 & 1.55 & 2.9\\
     J1232-4742 & 26.0 & 1.873 & 0.014 & 0.95 &  0.6\\
     J1503+2111 & 3.26 & 3.314 & 0.14 & 1.3 & 1.8\\
     J2251-3711 & 12.1 & 12.122 & 13.0 & 2.1 &  12.8\\
     \hline
                
    \end{tabular}
    \caption{Pulsars selected for this study. parameter $a_{vg}$ is the ratio of polar cap radius and the dimension of spark in vacuum gap model (GS00), while $a_{psg}$ is the ratio of spark area and polar cap area under PSG framework \citep{single_park}. The parameters $a_{vg}$ and $a_{psg}$ tell us how close these pulsars are to dying according to the single spark model in the vacuum gap and PSG framework, respectively. These parameters were computed using equation \ref{eqn:avg} and \ref{eqn:apsg}, with a reference that pulsar J2144-3933 (8.5 s pulsar) should have only a single spark.}
    \label{tab:source_list}
\end{table*}

\section{Similarity of death line pulsars with the general population}\label{sec:similarity}
Since pulsars in our sample belong to an extreme end of the $\mathrm{P} - \dot{\mathrm{P}}$ distribution, it is important to confirm that these pulsars are not very different from the normal pulsars and we can apply theoretical concepts developed for normal pulsars on these pulsars. We summarize the emission properties of the death-line pulsars in our sample based on the available literature and our uGMRT observations. We focus on a few well-established trends and relations in the current pulsar population. The aim is to establish the similarities of these death line pulsars with the general pulsar population.\\

\subsection{Duty-cycle period relation}
\begin{figure*}
    \centering
    \includegraphics[width=0.6\linewidth]{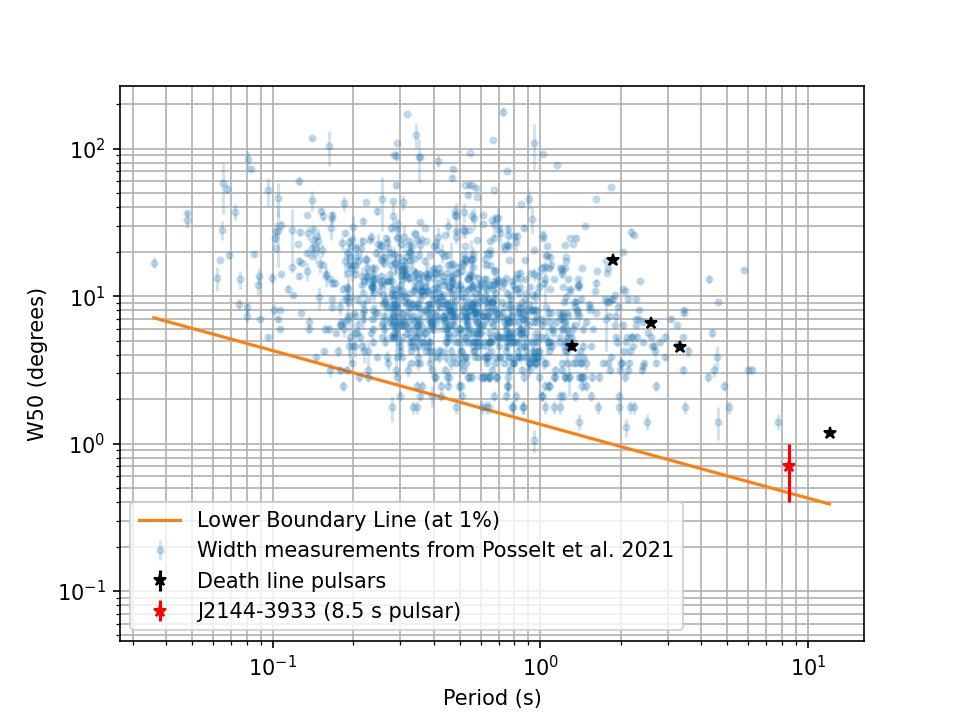}
    \caption{Period versus duty-cycle of death-line pulsars (block stars) along with a subset of the current population. Errors in the W50 measurements of death line pulsars are smaller than the size of the markers. The W50 measurement for pulsar J2144-3933 (red star) has been taken from \citet{EPN_1379}. The orange line is the lower boundary line on the duty cycle that allows $1\%$ of the population below it.}
    \label{fig:duty_period}
\end{figure*}

As per the current understanding, pulsar magnetic field geometry is dipolar at the heights where radio emission originates \citep{Mitra_2017}. The opening angle of the radio beam is expected to reduce for longer periods. Though the observed duty cycle of a pulsar depends on both the geometry of the pulsar and the opening angle of the radio beam, the reduction of the opening angle results in a trend of decreasing duty cycle with increasing period \citep{TPA_width, duty_vs_period_johnston}. Figure \ref{fig:duty_period} shows the duty-cycle (W50 measurements) of a subset of the current population at 1285 MHz central frequency \citep{TPA_width} along with the L-band measurement of the duty-cycle of the death-line pulsars in our sample. The pattern of decreasing duty cycle with increasing periods is followed by the death-line pulsars along with the general pulsar population. \\

\subsection{Radius to frequency mapping}
The standard pulsar emission models expect higher radio frequencies to be emitted closer to the stellar surface, giving rise to a relation called radius to frequency mapping \citep{Thorsett_1992, RFM_book}. This phenomenon is reflected in the profile width at different frequencies. Due to the dipolar geometry, the radio beams become wider as we move away from the stellar surface. This results in wider profile widths at lower frequencies and narrower profile widths at higher frequencies. This relation between profile width and observing frequency is followed by the majority of the pulsar population, though there are some exceptions \citep{RFM_examples}. We list the duty cycles of the five death line pulsars in our sample in table \ref{tab:sample_width}. Width at 50$\%$ height (W50) and width at 10$\%$ height (W10) are listed at two frequencies, 650 MHz (uGMRT observations) and 1400 MHz (from literature). The ratios of profile width at these two frequencies are shown in the R10 (ratio of W10 at frequencies 650 MHz and 1.4 GHz) and R50 (ratio of W50 at frequencies 650 MHz and 1.4 GHz) columns of the table. We see that 3 out of 5 pulsars show R50 values greater than or equal to 1.0, agreeing with the observed trend of wider profiles at lower frequencies. The W10 values should be a better measure of the actual radio beam size, but sometimes getting the width at 10$\%$ height becomes difficult due to low sensitivity. We could calculate R10 values  for only 3 pulsars, and all three agree with the width-to-frequency mapping trend. Overall, the death line pulsars largely (4 out of 5 pulsars in our sample) follow the trend of wider profiles at lower frequencies, as expected from the Radius to Frequency Model. Furthermore, some pulsars in our sample (J1115-0956, J1232-4742, and J1503+2111, see Fig \ref{fig:profiles}) show hints of increasing separation between components with decreasing frequency as seen by \citet{Thorsett_1992} in some other pulsars. \\

\begin{table*}
\centering
\begin{tabular}{c|c|c|c|c|c|c}
    Pulsar & W10 (650 MHz) & W10 (1.4 GHz) & R10 & W50 (650 MHz) & W50 (1.4 GHz) & R50\\
    \hline
    J0343-3000 & - & 28.8(7) & - & 4.9(7) & 5.3(2) \citep{TPA_width} & 0.9(2) \\
    J1115-0956 & 9.0(4) & - & - & 4.9(3) & 4.6(3) \citep{superb_v} & 1.0(1)\\
    J1232-4742 & 73(5) & 66(1) & 1.1(1) & 13.2(6) & 17.6(2) \citep{TPA_width} & 0.75(4)\\
    J1503+2111 & 9.8(3) & 8.1(3) & 1.2(1) & 4.9(7) & 4.5(2)\citep{TPA_width}  & 1.1(2)\\
    J2251-3711 & 2.9(3) & 2.14(3) & 1.3(1) & 1.3(1) & 1.18(1)\citep{morello_12s} & 1.10(7)\\
\end{tabular}
    \caption{Profile widths of pulsars in our sample. All the width measurements are in degrees. We measured the W10 (width at 10$\%$ height) and W50 (width at 50$\%$ height) at 650 MHz. The values of W10 and W50 at 1.4 GHz are taken from literature, the references are given in the W50(1.4 GHz) column. The numbers written in brackets are uncertainties in the last digit of the measurements. The R10 and R50 are the ratios of widths measured at frequencies 650 MHz and 1400 MHz for W10 and W50 respectively. The observations used in \citet{TPA_width} have a wide bandwidth of 775 MHz centered at 1284 MHz.}
    \label{tab:sample_width}
\end{table*}

\subsection{Rotating vector model}
Another signature of the dipolar magnetic field is seen in the form of the Rotating Vector Model (RVM, \citealt{RVM}) sweep of the polarization position angle (PPA). \citet{TPA_RVM1} recently classified a large sample of normal pulsars (854 pulsars) in two classes: first, where PPA sweep agrees with RVM (RVM class) and second where PPA sweep deviates significantly from RVM (non-RVM class). They estimate that 40$\%$ of the population belongs to non-RVM class. In a separate work, \citet{TPA_RVM2} show that RVM sweep can be recovered from the non-RVM class pulsars if only highly polarized time samples are used. The death line pulsar J0343-3000 was classified as RVM class by \citet{TPA_RVM1}, while J1232-4742 was classified as a flat class pulsar, which has RVM-like but flat PPA sweep. An RVM fitting on the death line pulsar J1503+2111 was not attempted by \citet{TPA_RVM1} due to low polarization and sensitivity issues. Though the position angle sweep of the folded profile has a complicated shape for pulsar J2251-3711, \citet{morello_12s} found an RVM-like PPA sweep in a few single pulses. The L band profile of pulsar J1115-0956 does not have enough linear polarization to determine its PPA sweep (see Fig A1 of \citealt{superb_v}). So, out of five death line pulsars, two pulsars don't have enough linear polarization to estimate the PPA sweep. The remaining three pulsars show the signature of RMV sweep in either folded profile or single pulses. Overall, like the general pulsar population, some of these pulsars in the pulsar death valley agree with the Rotating vector model. \\

\subsection{Microstructures}
\begin{figure*}
    \subfloat[J1503+2111]{
    \includegraphics[width=0.5\textwidth]{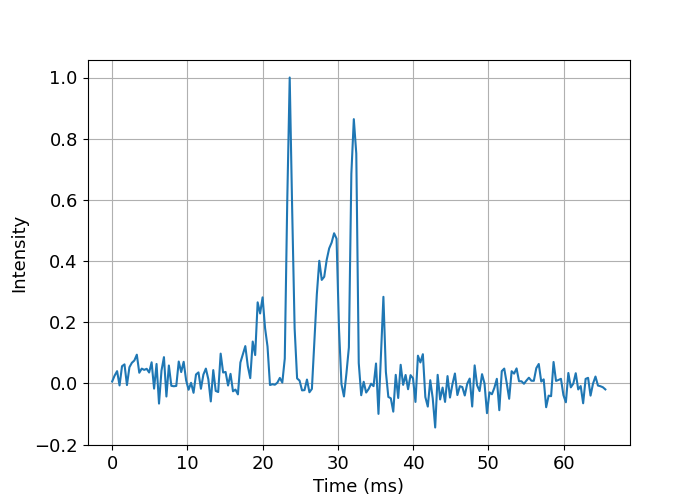}
    }
    \subfloat[J2251-3711]{
    \includegraphics[width=0.5\textwidth]{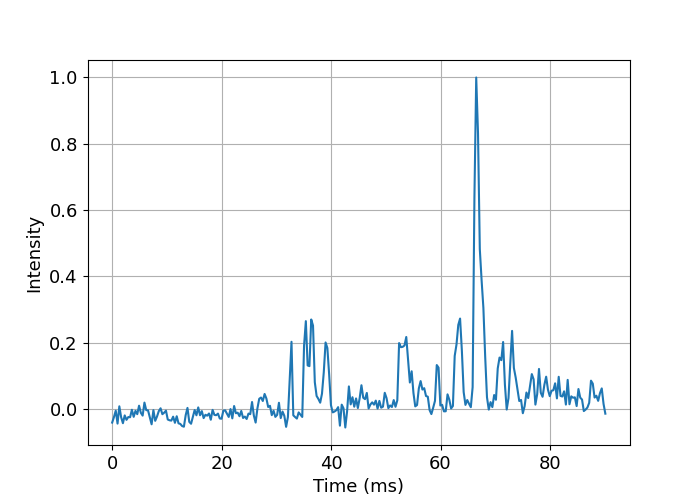}
    }
    \caption{Microstructures in two death line pulsars: J1503+2111 and J2251-3711.}
    \label{fig:micro-example}
\end{figure*}

\begin{figure*}
    \subfloat[J1503+2111]{
    \includegraphics[width=0.5\textwidth]{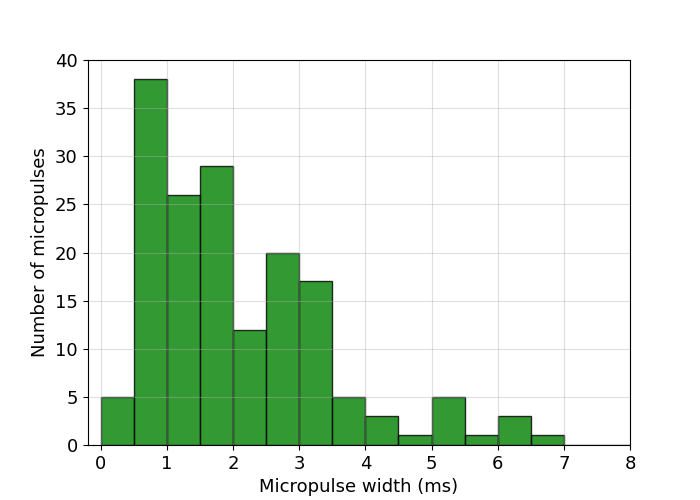}
    }
    \subfloat[J2251-3711]{
    \includegraphics[width=0.5\textwidth]{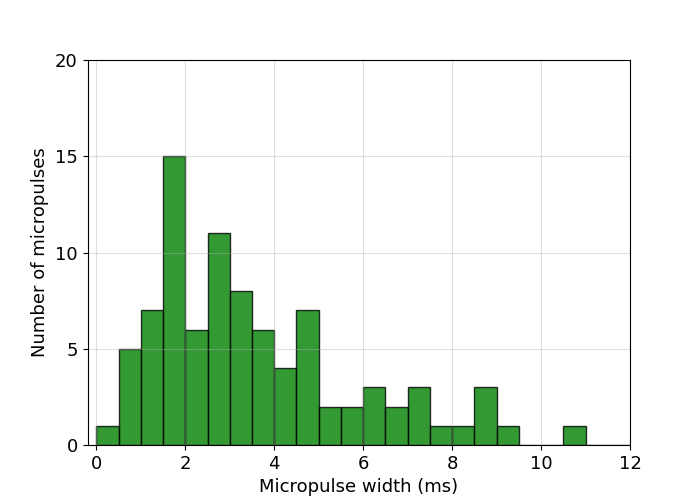}
    }
    \caption{Distribution of microstructure timescale in pulsars J1503+2111 and J2251-3711. We estimate a mean timescale of 2.0$\pm$0.3 for pulsar J1503+2111 using 166 individual micropulses and a timescale of 3.8$\pm$0.3 for pulsar J2251-3711 using 91 individual micropulses.}
    \label{fig:micro-width}
\end{figure*}

Microstructures are defined as narrow quasiperiodic structures seen in single pulses of some pulsars. We spotted narrow, occasionally quasi-periodic structures in single pulses of two death line pulsars, pulsars J1503+2111 and J2251-3711. Since these pulsars have very narrow duty cycles we do not see proper micropulse train in most of the single pulses, but spike-like structures with timescales much smaller than the on-pulse window are commonly seen (see Fig. \ref{fig:micro-example}). To measure the timescales of these spikes, we apply the method used by \citet{intrinsic_micro} to measure timescales of intrinsic micropulses. We collected 166 micropulses from pulsar J1503+2111 and 91 micropulses from pulsar J2251-3711. Figure \ref{fig:micro-width} shows the micropulse width distribution of these pulsars. The mean micropulse width is 2.0$\pm$0.3 for pulsar J1503+2111 and 3.8$\pm$0.3 for pulsar J2251-3711. These measurements roughly agree with the known relation between microstructure width and the period of pulsars \citep{micro_kramer}.

\subsection{Subpulse drifting and change of emission modes}\label{subsec:mode_and_drift}
\begin{figure}
    \centering
    \includegraphics[width=\linewidth]{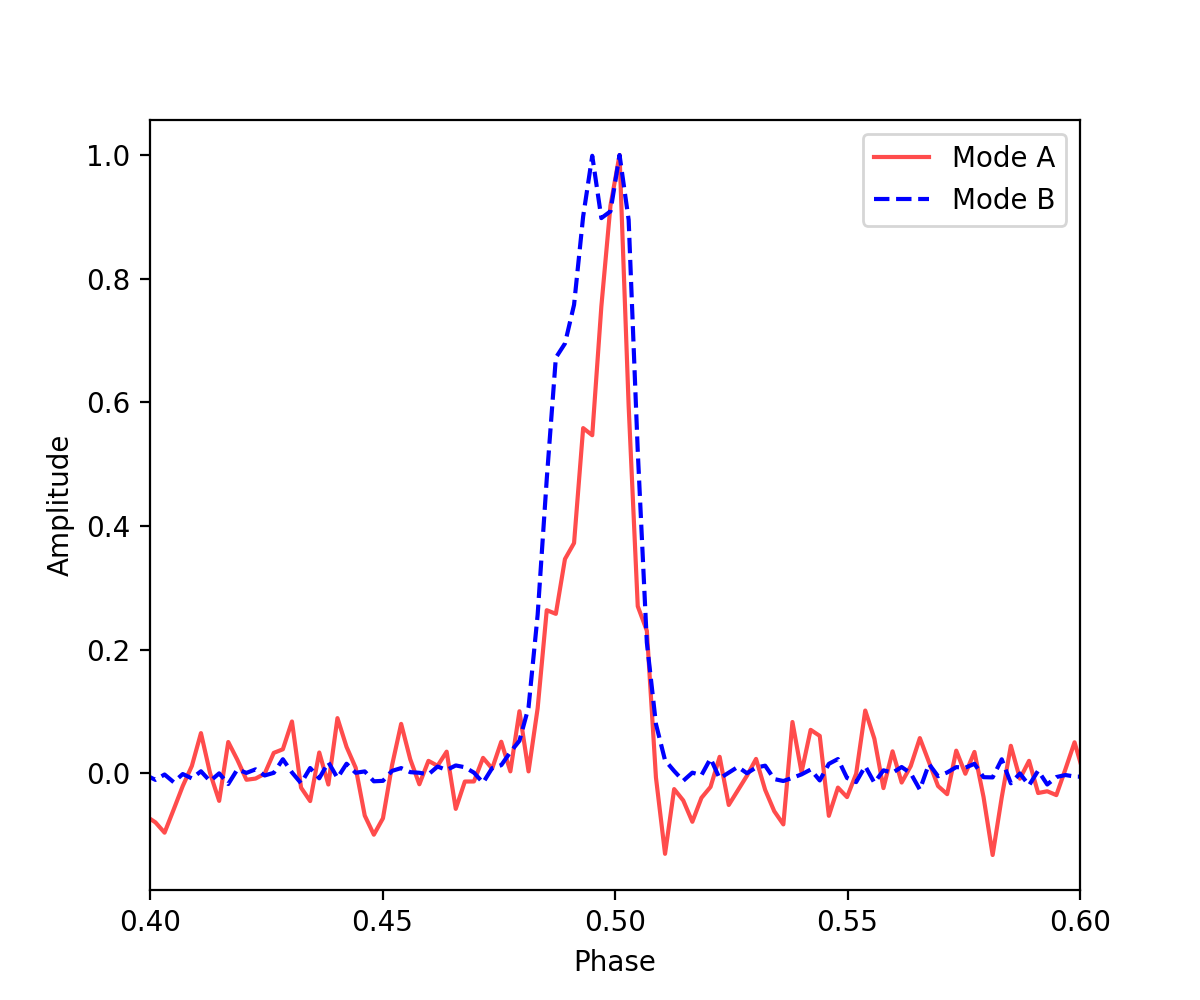}
    \caption{Two emission modes of pulsar J1503+2111. Mode A profile (red profile) was generated by folding a 52-minute-long observation on 14 Nov 2023. Mode B (dashed blue profile) was generated by folding a 55-minute-long observation on 04 Feb 2022.}
    \label{fig:J1503_modes}
\end{figure}

 Two out of five death line pulsars in our sample are known to show subpulse drifting. \citet{TPA_drifting} characterized the subpulse drifting and determined the drifting periodicity (P3) for these pulsars. The P3 timescale for pulsar J1232-4742 is 19.6$\pm$0.2 periods and the P2 value (separation between two subpulses) is around 20 degrees \citep{TPA_drifting}. As the W10 value of this pulsar is 66 degrees at 1400 MHz, three subpulses can fit within the on-pulse window. Hence, this pulsar seems to have at least three sparks (probably many more) on its polar cap. For the pulsar J1503+2111, the P3 value is 16$\pm$2 periods \citep{TPA_drifting}. We also encounter very different folded profile shapes for pulsar J1503+2111 in two different uGMRT observations (see Fig. \ref{fig:J1503_modes}). The two profiles are stable as they have been obtained after folding enough rotation cycles of the pulsar. The profile obtained from the observation on 14 Nov 2023 (mode A) resembles the profile shape available in the literature for this pulsar (see Fig \ref{fig:profiles} (d)). The profile from the 04 Feb 2022 observation (mode B) is the new emission mode of this pulsar discovered in this work. The beam subtraction (PA-IA) performed on 04 Feb 2023 observation has greatly reduced the noise level in the profile.

\section{Evidence for multiple sparks in death line pulsars}\label{sec:profiles}
\citet{single_park} noted that a pulsar with only a single spark on its polar cap will have certain emission properties. They highlighted that such a pulsar can only have a single component profile. Pulsars with only a single spark on the polar cap can not show certain emission features like subpulse drifting and mode changing. Similarly, GS00 also showed that the complexity parameter ($a_{vg}$) is the upper limit on the number of components in the pulsar profile (i.e. profile components $\le a_{vg}$). Hence, pulsars with $a_{vg}=1$ can have only a single component profile. If we consider that sparks on the polar cap have a dimension $h$ and are also separated by a distance h (see Figure 1 of GS00), a pulsar should have $a_{vg} >= 3$ to have at least one cone in the radio beam and therefore more than one profile component. Now, we will go through the profiles of the individual death-line pulsars in our sample.

\begin{figure*}
    \subfloat[J0343-3000]{
    \includegraphics[width=0.31\textwidth]{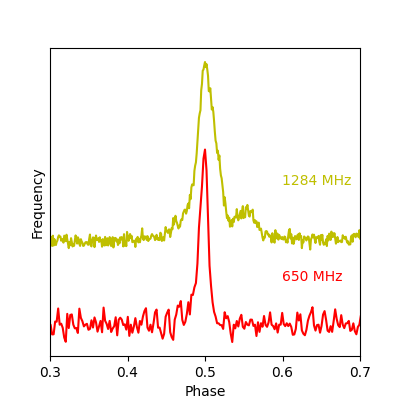}
    }
    \subfloat[J1115-0956]{
    \includegraphics[width=0.31\textwidth]{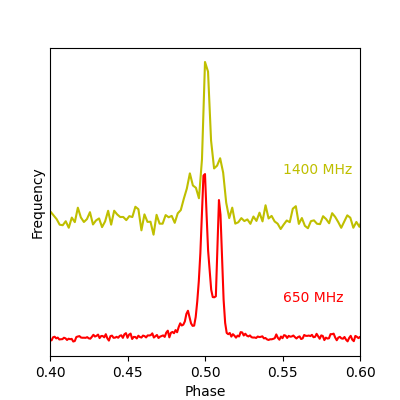}
    }
    \subfloat[J1232-4742]{
    \includegraphics[width=0.31\textwidth]{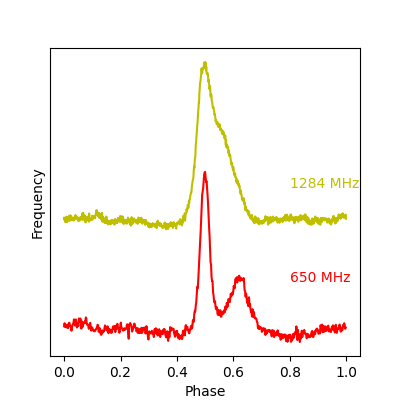}
    }    
    \hspace{00mm}
    \subfloat[J1503+2111]{
    \includegraphics[width=0.35\textwidth]{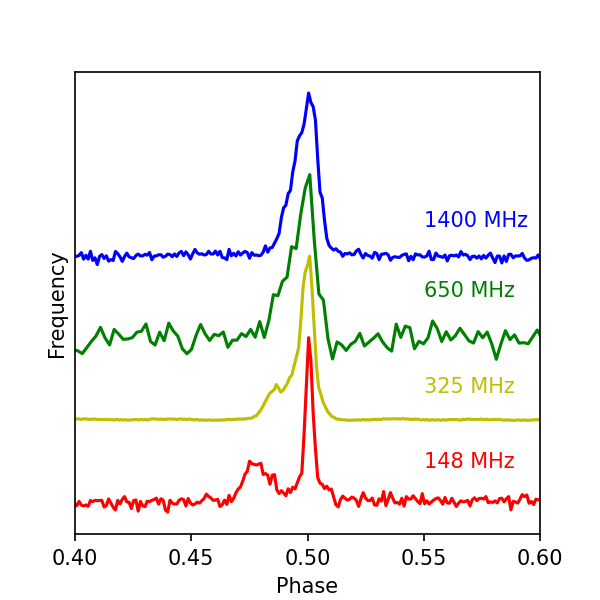}
    }
    \subfloat[J2251-3711]{
    \includegraphics[width=0.35\textwidth]{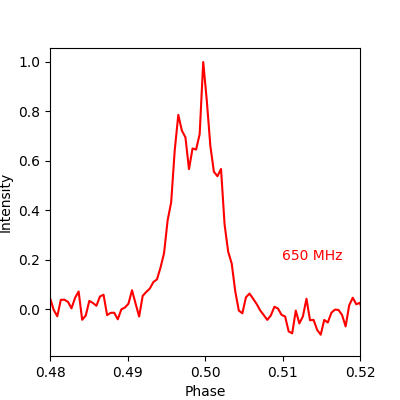}
    }
    \caption{Profiles of five death line pulsars. We have plotted publicly available profiles of these pulsars along with profiles we detect with the uGMRT (profiles corresponding to 650 MHz central frequency). We got profiles of four pulsars at least at two widely separated frequencies. These four pulsars show significant profile evolution with frequency. All five pulsars show clear multicomponent profiles at some frequencies.}
    \label{fig:profiles}
\end{figure*}

The Figure \ref{fig:profiles} shows profiles of the five Death Valley pulsars collected from the literature, along with the profiles from the uGMRT observations. All the profiles at 650 MHz shown in these plots are from the uGMRT observations conducted in this work. Figure \ref{fig:profiles}(a) shows the folded profile of pulsar J0343-3000 at central frequencies 1285 MHz \citep{TPA_width} and 650 MHz. The folded profile of this pulsar shows two components at 1285 MHz. The folded profile of pulsar J1115-0956 (see Fig. \ref{fig:profiles}(b)) has three components visible at both 1400 MHz \citep{superb_v} and 650 MHz central frequencies. Figure \ref{fig:profiles}(c) shows the folded profiles of pulsar J1232-4742 at central frequencies 1284 MHz  \citep{TPA_width} and 650 MHz. This pulsar clearly shows two components at 650 MHz. Figure \ref{fig:profiles}(d) shows the folded profile of pulsar J1503+2111 at four different central frequencies. The profile at 148 MHz has been taken from \citet{J1503_lofar}, the profile at 325 MHz and 1400 MHz are from \citet{J1503_325MHz}. This pulsar shows two clear components in the profile at 148 MHz. The pulsar J2251-3711 does not show a clear multicomponent profile in L-band \citep{morello_12s}. Our observation in the uGMRT band-4 (550-750 MHz) reveals an additional component in the profile of this pulsar. Fig. \ref{fig:profiles}(e) shows the profile of the pulsar from our observation on 04 Feb 2022. The observation duration is only 1.1 hours, which accommodates $\sim 340$ pulses from this pulsar.

Summarizing the figure \ref{fig:profiles}, all five pulsars in our sample show multi-component profiles at some frequency. According to GS00 a multi-component profile requires $a_{vg} >=3$. Most of the pulsars in our sample (except for J2251-3711) have $a_{vg} < 2$ but still show multi-component profiles. The partially screened voltage gap (PSG) model predicts less than two sparks ($a_{psg} < 2$) for pulsars J0343-3000, J1232-4742, and J1503+2111, even though these pulsars show multi-component profiles. Pulsar J1232-4742, with $a_{vg} = 0.97$ and $a_{psg} = 0.6$, should not emit in radio wavelengths. However, this pulsar shows a two-component profile. Pulsar J1503+2111 and J1232-4742 show further evidence of multiple sparks on their polar cap, in the form of subpulse drifting and emission mode change.

\begin{figure*}
    \centering
    \includegraphics[width=0.7\textwidth]{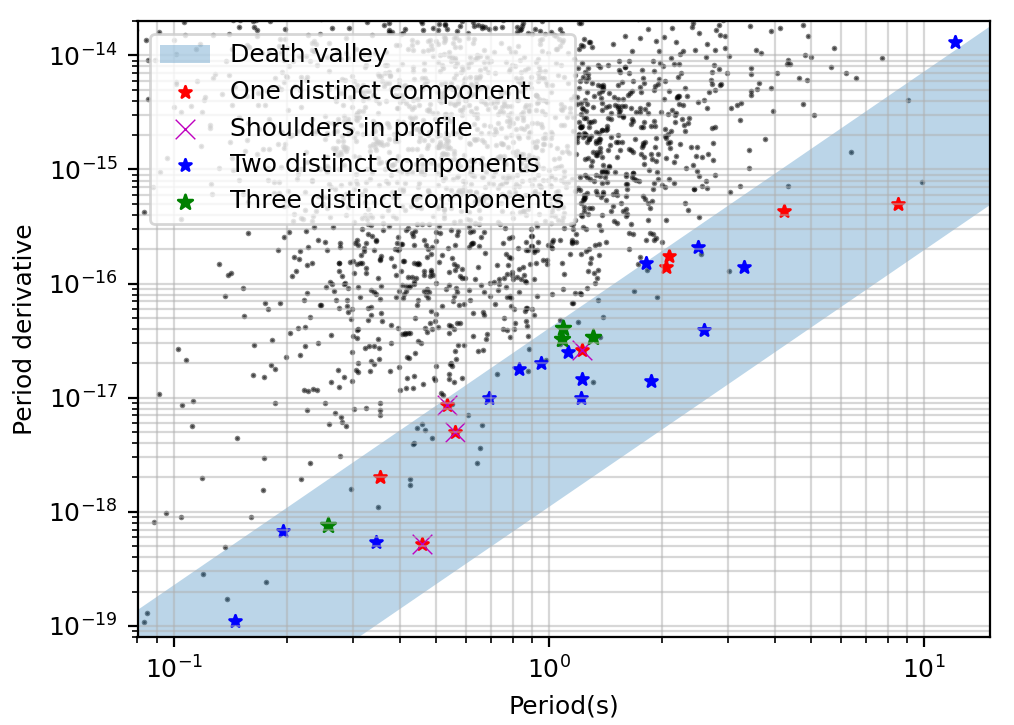}
    \caption{Pulsars in the single spark death valley of the vacuum voltage gap model. Out of these 28 pulsars, 4 show 3 distinct components, 15 show two distinct components, while 9 show a single component profile. Out of these 9 single component profiles, 4 profiles (marked with '$\times$') show shoulders (breaks in the slope), hinting towards possibility of multiple merged components.}
    \label{fig:vacuum_death_valley}
\end{figure*}

\begin{figure*}
    \centering\includegraphics[width=\linewidth]{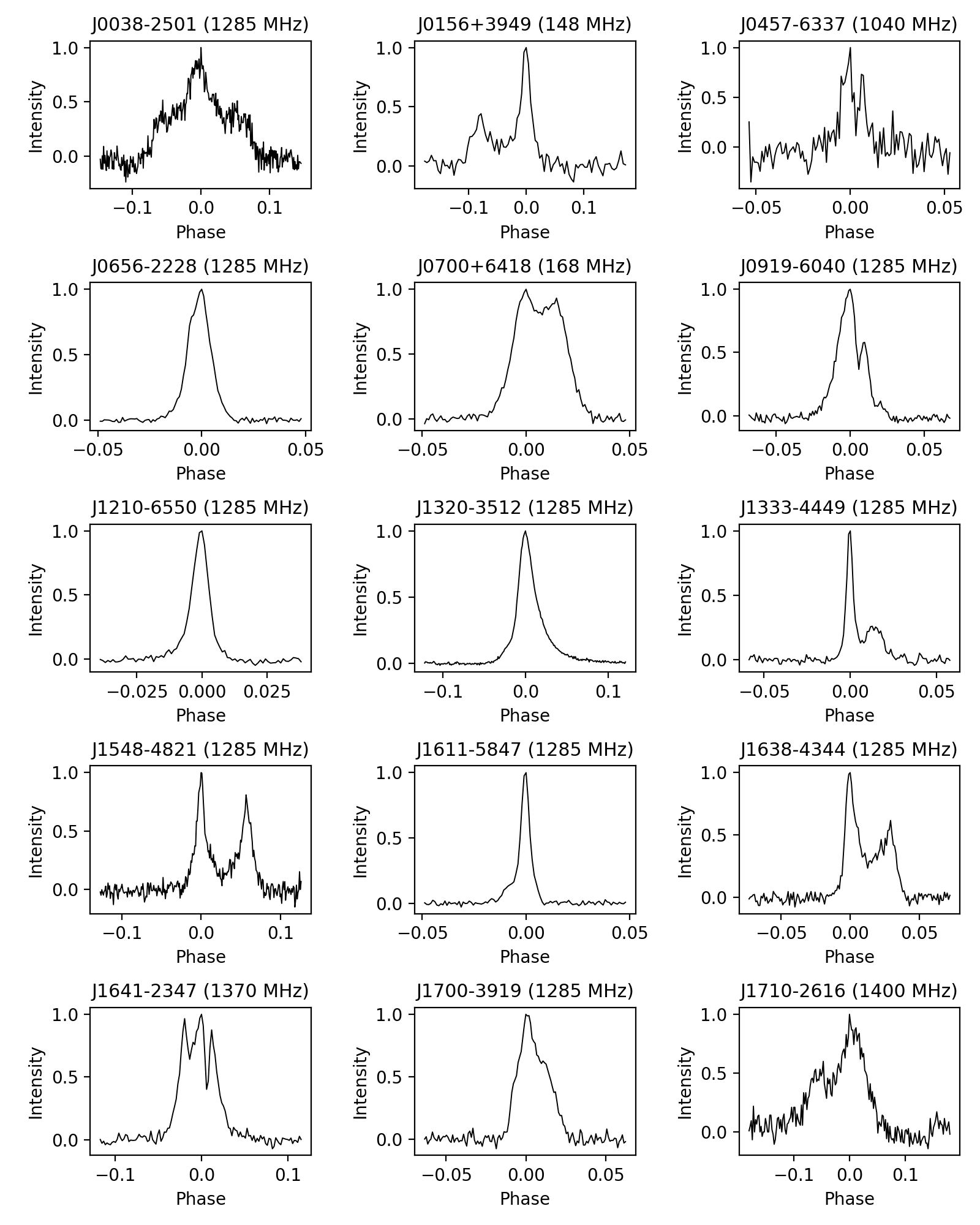}
    \caption{Profile of 23 pulsars in the vacuum voltage gap single spark death valley. The profiles have been taken from the available literature (see table \ref{tab:additional_pulsars} for references).}
    \label{fig:additional_profiles}
\end{figure*}

Even though all five pulsars in our sample disagree with the single spark model of pulsar death, the sample size is too small to draw any reliable conclusion. To increase the sample size, we shortlisted pulsars with periods greater than 100 ms in the single spark death valley of the vacuum voltage gap model (see Fig. \ref{fig:vacuum_death_valley}) and searched for their folded profile in the available literature, mainly in the EPN database of pulsar profiles \footnote{\url{https://psrweb.jb.man.ac.uk/epndb/}} and the publically available Thousand Pulsar Array (TPA) data\footnote{\url{https://zenodo.org/records/7272361}} \citep{TPA_datarelease}. We found profiles for 23 pulsars in the single spark death valley (excluding the pulsars already included in this work). These pulsars, along with the corresponding parameters $a_{vg}$, $a_{psg}$, and the number of distinct components in their profile, are listed in Table \ref{tab:additional_pulsars}. The profiles of these 23 pulsars are shown in Fig. \ref{fig:additional_profiles}. If profiles at more than one frequency are available for a pulsar, we take the profile with the maximum number of distinct components.

Figure \ref{fig:vacuum_death_valley} shows the location of all 28 pulsars (including five pulsars from our initial sample) on the $\mathrm{P}-\dot{\mathrm{P}}$ plane. The number of distinct components in the profiles are indicated by different colors. Four pulsars ($14\%$ of the sample) show three components, 15 pulsars ($54\%$ of the sample) show two distinct components, while 9 pulsars ($32\%$ of the sample) show only a single component in their folded profiles. 4 out of 9 single-component profile pulsars show shoulders in their profile (J0656-2228, J1611-5847, J1700-3919, and J2307+2225, marked by `$\times$' in the Fig. \ref{fig:vacuum_death_valley}), potentially indicating multiple merged components. There is a chance that these shoulders will turn out to be separate components in a multi-frequency observation. Overall, only 9 out of 28 ($\sim 32\%$) pulsars in the single spark death valley potentially agree with the single spark model of pulsar death. Multi-frequency and polarization studies of these 9 pulsars are required to confirm if they really agree with the single spark model of pulsar death. A majority of pulsars in the single spark death valley (19 of 28 pulsars, $\sim 68\%$) show multi-component profiles, disagreeing with the single spark model of pulsar death.

\begin{table*}
    \centering
    \begin{tabular}{lccccccr}
      Pulsar name & Period & Period-derivative & $a_{vg}$ & $a_{psg}$ & Number of components & Reference for profile \\
                 & (s) & ($\times 10^{-15}~s/s$) & ($r_p/h$, vacuum gap) & ($A_{pc}/A_{sp}$, PSG) & in profile & \\
                \hline
     J0038-2501 & 0.2569 & 0.00076 & 1.5 & 1.7 & 3 & \citet{TPA_datarelease}\\ 
     J0156+3949 & 1.811 & 0.151 & 1.9 &  6.7 & 2 &  \citet{J1503_lofar}\\
     J0457-6337 & 2.497 & 0.21 & 1.7 & 4.9 & 2 & \citep{TPA_datarelease} \\
     J0656-2228 & 1.224 & 0.026 & 1.5 & 2.6 & 1 & \citet{TPA_datarelease} \\
     J0700+6418 & 0.1956 & 0.00068 & 1.7 & 2.6 & 2 & \citet{J1503_lofar} \\
     J0919-6040 & 1.216 & 0.01 & 1.1 & 0.98 & 2 & \citet{TPA_datarelease} \\
     J1210-6550 & 4.237 & 0.43 & 1.5 & 3.5 & 1 & \citet{TPA_datarelease} \\
     J1320-3512 & 0.4584 & 0.00052 & 0.92 & 0.36 & 1 & \citet{TPA_datarelease} \\
     J1333-4449 & 0.3456 & 0.00054 & 1.1 & 0.7 & 2 & \citet{TPA_datarelease} \\
     J1548-4821 & 0.1456 & 0.0001 & 1.2 & 0.8 & 2 & \citet{TPA_datarelease} \\
     J1611-5847 & 0.3545 & 0.002 & 1.6 & 2.3 & 1 & \citet{TPA_datarelease} \\
     J1638-4344 & 1.121 & 0.025 & 1.5 & 2.9 & 2 &  \citet{TPA_datarelease} \\
     J1641-2347 & 1.091 & 0.041 & 1.8 & 5.0 & 3 & \citet{EPN_1379} \\
     J1700-3919 & 0.5605 & 0.005 & 1.5 & 2.3 & 1 &  \citet{TPA_datarelease} \\
     J1710-2616 & 0.9541 & 0.02 & 1.6 & 3.2 & 2 & \citet{EPN_1379} \\
     J1714-1054 & 2.088 & 0.17 & 1.8 & 5.9 & 1 & \citet{EPN_1379} \\
     J1717-3953 & 1.085 & 0.032 & 1.7 & 4.0 & 3 & \citet{EPN_1379} \\
     J1901+1306 & 1.830 & 0.131 & 1.8 & 5.7 & 2 & \citet{J1901_prof} \\
     J1914+0631 & 0.6938 & 0.0099 & 1.8 & 4.0 & 2 & \citet{TPA_datarelease} \\
     J1915+0752 & 2.058 & 0.139 & 1.7 & 4.8 & 1 & \citet{TPA_datarelease} \\
     J2136-1606 & 1.227 & 0.014 & 1.2 & 1.4 & 2 & \citet{TPA_datarelease} \\
     J2144-3933 & 8.509 & 0.496 & 1.0 & 1.0 & 1 & \citet{TPA_datarelease} \\
     J2307+2225 & 0.5358 & 0.0087 & 1.8 & 4.4 & 1 & \citet{TPA_datarelease}\\
     \hline
                
    \end{tabular}
    \caption{More pulsars in the vacuum voltage gap death valley. The parameters ($a_{vg}$ and $a_{psg}$) were computed using equation \ref{eqn:avg} and \ref{eqn:apsg}, with a reference that pulsar J2144-3933 (8.5 s pulsar) should have only a single spark. The number of components can be estimated from the folded profiles provided in Figure \ref{fig:additional_profiles}. The last column lists the references for the profile of the respective pulsars.}
    \label{tab:additional_pulsars}
\end{table*}

\renewcommand{\thefigure}{\arabic{figure} (Continued.)}
\addtocounter{figure}{-1}
\begin{figure*}
    \centering\includegraphics[width=\linewidth]{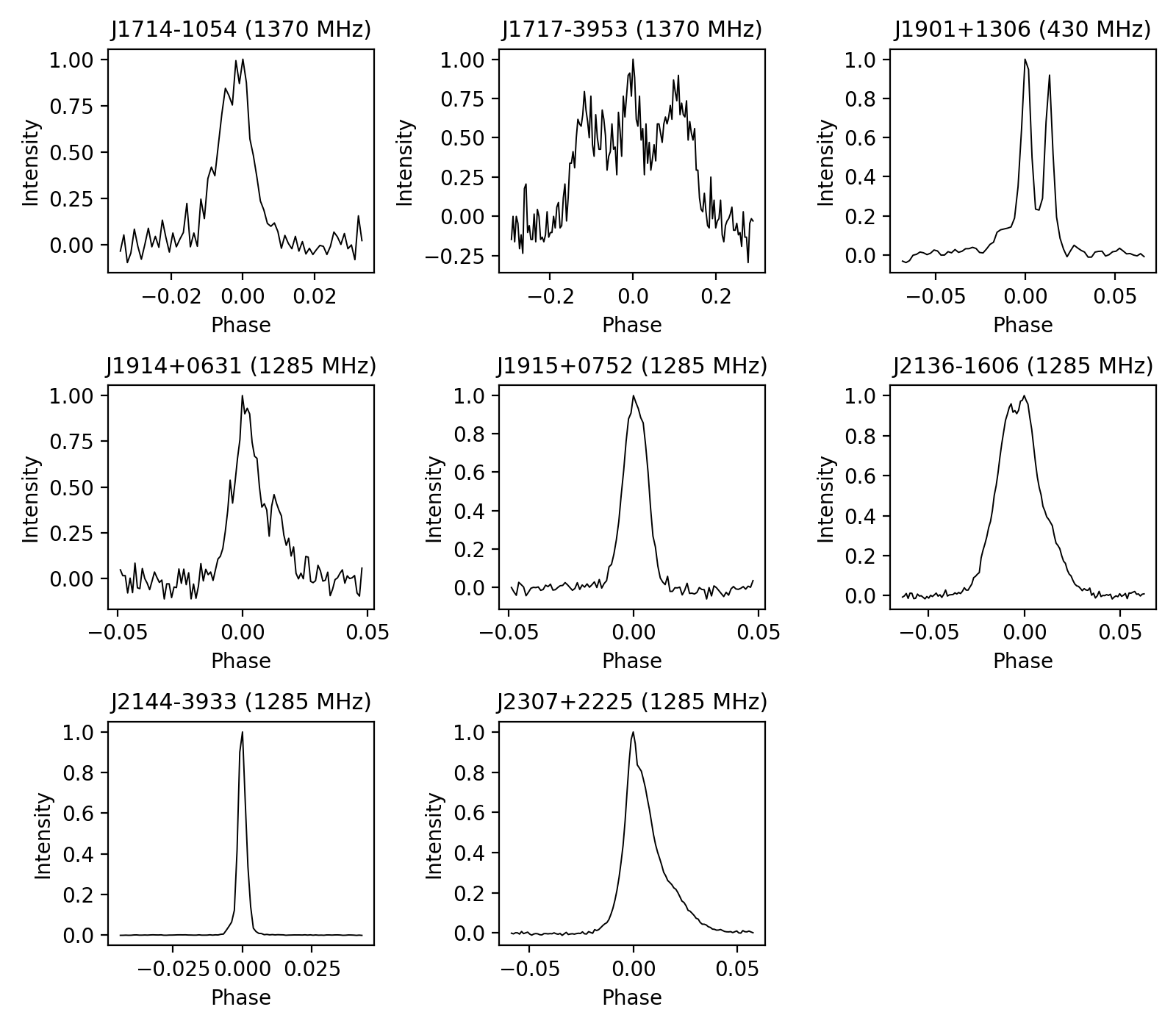}
    \caption{}
\end{figure*}
\section{Discussion}\label{sec:discussion}
The death of a radio pulsar is caused by the lack of charged particles that produce radio emission through curvature radiation. These charged particles are produced in the pair cascade process in a voltage gap on the polar cap. The pair cascade process requires a minimum amount of voltage difference to accelerate the charged particles and produce energetic photons, and then sufficiently curved magnetic field lines to support pair production. Now, the physical conditions on the polar cap depend on what voltage gap model is assumed. There are two popular voltage gap models: the classical vacuum voltage gap model (RS75) and its modified version called the partially screened gap model \citep{Gil_2003}. \citet{single_park} predict only a single spark on the polar cap of a dying pulsar (a pulsar that is at the brink of ending its radio emission) under the framework of the partially screened vacuum gap model. A closer look at the vacuum gap death line model \citep{chen_and_rutherman} reveals that only a single spark is available on the polar cap in the vacuum gap framework as well. We expect some particular emission properties from a pulsar that has only a single spark on its polar cap. Such pulsars can not show subpulse drifting and mode changing, but most importantly, a multi-component profile \citep{single_park}.

We studied the emission properties of five death line pulsars in our sample. These pulsars have emission properties very similar to the normal pulsar population. The majority of these pulsars agree with the well-established trends like radius to frequency mapping \citep{RFM_book}, rotating vector model (RVM, \citealt{RVM}), and trend of decreasing duty-cycle with increasing rotation period \citep{TPA_width, LBL_interpulse}. Two of these pulsars show microstructures and the microstructure timescales roughly agree with the known relation between microstructure timescale and pulsar period \citep{Kramer_2002}. Though subpulse drifting and emission mode change are not expected from pulsars with a single spark \citep{single_park}, two of these pulsars are known to show subpulse drifting. We also discovered emission mode change phenomena in the pulsar J1503+2111.

 We collected folded profiles of 23 more pulsars in the death valley from the available literature, which, combined with the five pulsars we observed with the uGMRT, gives a set of 28 profiles of death valley pulsars. We found that majority of these pulsars (19 out of 28, $\sim 68\%$) show multi-component profiles, disagreeing with the single spark model of pulsar death. Only 9 out of these 28 pulsars ($\sim 32\%$) show single-component profiles, potentially agreeing with the single spark pulsar death model. Multi-frequency and polarization studies of these pulsars are required to show if they support the presence of a single spark on the polar cap. Two pulsars, J1320-3512 and J1232-4742, in the sample of 28 death valley pulsars, offer very constraining cases for the single spark models of pulsar death. These pulsars should not be emitting in radio frequencies under both the voltage gap frameworks, as both $a_{vg}$ and $a_{psg}$ are less than one. To explain the two-component profile of pulsar J1232-4742 in the vacuum voltage gap model, we need $a_{vg}> 3$, but this is very unlikely considering the very weak dependence of $a_{vg}$ on the surface magnetic field configuration. The PSG death line model has a strong dependence on the surface magnetic field configuration, but still, this pulsar will require an unrealistic multipolar and curved surface field configuration. This highlights the inconsistency between what is expected according to theoretical models for pulsar death and what is seen in the pulsars that are actually at the lower boundary of the  $\mathrm{P}-\dot{\mathrm{P}}$ plane.

 In the past few years, there have been several discoveries of Ultra-Long Period (ULP) sources with periods greater than 1 minute (e.g. \citealt{ULP_76s}, \citealt{ULP_421s}, \citealt{ULP_44m}, \citealt{ULP_54m}, and \citealt{ULP_2.9h}). Some of these sources are likely to be neutron stars \citep{NS_ULP}. Due to their very slow rotation, many of these sources challenge the death line models for radio-emitting neutron stars \citep{ULP_summary}. Recent population synthesis studies, accounting for the magnetic axis alignment and magnetic field decay with age, suggest that the current pulsar population can be easily reproduced without requiring any pulsar death phenomena \citep{population2, population1}. All these new developments demand a careful reconsideration of the pulsar death phenomena in the pulsar radio emission models. Also, the alternative models for pulsar radio emission, currently being developed, such as particle-in-cell simulations \citep{PIC_2024, pic_2023, PIC_2020, PIC_2010}, might be able to address these concerns regarding the pulsar death phenomenon.

 \textbf{}

\section*{Acknowledgements}
 The GMRT is run by the National Centre for Radio Astrophysics of the Tata Institute of Fundamental Research, India. We acknowledge the support of GMRT telescope operators and the GMRT staff for supporting the observations used in this study.

\section*{Data Availability}
The profiles for the five pulsars observed with the uGMRT will be uploaded to the EPN pulsar profile database. The data for other emission properties of these pulsars will be shared on reasonable request to the corresponding author.



\bibliographystyle{mnras}
\bibliography{example} 

\begin{thebibliography}{}
\makeatletter
\relax
\def\mn@urlcharsother{\let\do\@makeother \do\$\do\&\do\#\do\^\do\_\do\%\do\~}
\def\mn@doi{\begingroup\mn@urlcharsother \@ifnextchar [ {\mn@doi@}
  {\mn@doi@[]}}
\def\mn@doi@[#1]#2{\def\@tempa{#1}\ifx\@tempa\@empty \href
  {http://dx.doi.org/#2} {doi:#2}\else \href {http://dx.doi.org/#2} {#1}\fi
  \endgroup}
\def\mn@eprint#1#2{\mn@eprint@#1:#2::\@nil}
\def\mn@eprint@arXiv#1{\href {http://arxiv.org/abs/#1} {{\tt arXiv:#1}}}
\def\mn@eprint@dblp#1{\href {http://dblp.uni-trier.de/rec/bibtex/#1.xml}
  {dblp:#1}}
\def\mn@eprint@#1:#2:#3:#4\@nil{\def\@tempa {#1}\def\@tempb {#2}\def\@tempc
  {#3}\ifx \@tempc \@empty \let \@tempc \@tempb \let \@tempb \@tempa \fi \ifx
  \@tempb \@empty \def\@tempb {arXiv}\fi \@ifundefined
  {mn@eprint@\@tempb}{\@tempb:\@tempc}{\expandafter \expandafter \csname
  mn@eprint@\@tempb\endcsname \expandafter{\@tempc}}}

\bibitem[\protect\citeauthoryear{{Ben{\'a}{\v{c}}ek}, {Mu{\~n}oz},
  {B{\"u}chner}  \& {Jessner}}{{Ben{\'a}{\v{c}}ek} et~al.}{2023}]{pic_2023}
{Ben{\'a}{\v{c}}ek} J.,  {Mu{\~n}oz} P.~A.,  {B{\"u}chner} J.,   {Jessner} A.,
  2023, \mn@doi [\aap] {10.1051/0004-6361/202345987}, \href
  {https://ui.adsabs.harvard.edu/abs/2023A&A...675A..42B} {675, A42}

\bibitem[\protect\citeauthoryear{{Ben{\'a}{\v{c}}ek}, {Timokhin}, {Mu{\~n}oz},
  {Jessner}, {Rievajov{\'a}}, {Pohl}  \& {B{\"u}chner}}{{Ben{\'a}{\v{c}}ek}
  et~al.}{2024}]{PIC_2024}
{Ben{\'a}{\v{c}}ek} J.,  {Timokhin} A.,  {Mu{\~n}oz} P.~A.,  {Jessner} A.,
  {Rievajov{\'a}} T.,  {Pohl} M.,   {B{\"u}chner} J.,  2024, \mn@doi [\aap]
  {10.1051/0004-6361/202450949}, \href
  {https://ui.adsabs.harvard.edu/abs/2024A&A...691A.137B} {691, A137}

\bibitem[\protect\citeauthoryear{Beskin \& Istomin}{Beskin \&
  Istomin}{2022}]{deathline_beskin2}
Beskin V.~S.,  Istomin A.~Y.,  2022, \mn@doi [Monthly Notices of the Royal
  Astronomical Society] {10.1093/mnras/stac2423}, 516, 5084

\bibitem[\protect\citeauthoryear{{Beskin} \& {Litvinov}}{{Beskin} \&
  {Litvinov}}{2022}]{deathline_beskin1}
{Beskin} V.~S.,  {Litvinov} P.~E.,  2022, \mn@doi [\mnras]
  {10.1093/mnras/stab3575}, \href
  {https://ui.adsabs.harvard.edu/abs/2022MNRAS.510.2572B} {510, 2572}

\bibitem[\protect\citeauthoryear{{Bilous} et~al.,}{{Bilous}
  et~al.}{2016}]{J1503_lofar}
{Bilous} A.~V.,  et~al., 2016, \mn@doi [\aap] {10.1051/0004-6361/201527702},
  \href {https://ui.adsabs.harvard.edu/abs/2016A&A...591A.134B} {591, A134}

\bibitem[\protect\citeauthoryear{{Buch}, {Kale}, {Muley}, {Kudale}  \&
  {Ajithkumar}}{{Buch} et~al.}{2023}]{realtime_RFI}
{Buch} K.~D.,  {Kale} R.,  {Muley} M.,  {Kudale} S.,   {Ajithkumar} B.,  2023,
  \mn@doi [Journal of Astrophysics and Astronomy] {10.1007/s12036-023-09919-x},
  \href {https://ui.adsabs.harvard.edu/abs/2023JApA...44...37B} {44, 37}

\bibitem[\protect\citeauthoryear{{Caleb} et~al.,}{{Caleb}
  et~al.}{2022}]{ULP_76s}
{Caleb} M.,  et~al., 2022, \mn@doi [Nature Astronomy]
  {10.1038/s41550-022-01688-x}, \href
  {https://ui.adsabs.harvard.edu/abs/2022NatAs...6..828C} {6, 828}

\bibitem[\protect\citeauthoryear{{Caleb} et~al.,}{{Caleb}
  et~al.}{2024}]{ULP_54m}
{Caleb} M.,  et~al., 2024, \mn@doi [Nature Astronomy]
  {10.1038/s41550-024-02277-w}, \href
  {https://ui.adsabs.harvard.edu/abs/2024NatAs...8.1159C} {8, 1159}

\bibitem[\protect\citeauthoryear{{Camilo} \& {Nice}}{{Camilo} \&
  {Nice}}{1995}]{J1901_prof}
{Camilo} F.,  {Nice} D.~J.,  1995, \mn@doi [\apj] {10.1086/175737}, \href
  {https://ui.adsabs.harvard.edu/abs/1995ApJ...445..756C} {445, 756}

\bibitem[\protect\citeauthoryear{{Chen} \& {Ruderman}}{{Chen} \&
  {Ruderman}}{1993}]{chen_and_rutherman}
{Chen} K.,  {Ruderman} M.,  1993, \mn@doi [\apj] {10.1086/172129}, \href
  {https://ui.adsabs.harvard.edu/abs/1993ApJ...402..264C} {402, 264}

\bibitem[\protect\citeauthoryear{{Chen} \& {Wang}}{{Chen} \&
  {Wang}}{2014}]{RFM_examples}
{Chen} J.~L.,  {Wang} H.~G.,  2014, \mn@doi [\apjs]
  {10.1088/0067-0049/215/1/11}, \href
  {https://ui.adsabs.harvard.edu/abs/2014ApJS..215...11C} {215, 11}

\bibitem[\protect\citeauthoryear{{Coti Zelati} \& {Borghese}}{{Coti Zelati} \&
  {Borghese}}{2024}]{ULP_summary}
{Coti Zelati} F.,  {Borghese} A.,  2024, \mn@doi [arXiv e-prints]
  {10.48550/arXiv.2412.12763}, \href
  {https://ui.adsabs.harvard.edu/abs/2024arXiv241212763C} {p. arXiv:2412.12763}

\bibitem[\protect\citeauthoryear{{Dirson}, {P{\'e}tri}  \& {Mitra}}{{Dirson}
  et~al.}{2022}]{population1}
{Dirson} L.,  {P{\'e}tri} J.,   {Mitra} D.,  2022, \mn@doi [\aap]
  {10.1051/0004-6361/202243305}, \href
  {https://ui.adsabs.harvard.edu/abs/2022A&A...667A..82D} {667, A82}

\bibitem[\protect\citeauthoryear{{Dong} et~al.,}{{Dong}
  et~al.}{2024}]{ULP_421s}
{Dong} F.~A.,  et~al., 2024, \mn@doi [arXiv e-prints]
  {10.48550/arXiv.2407.07480}, \href
  {https://ui.adsabs.harvard.edu/abs/2024arXiv240707480D} {p. arXiv:2407.07480}

\bibitem[\protect\citeauthoryear{Gil \& Sendyk}{Gil \& Sendyk}{2000}]{Gil_2000}
Gil J.~A.,  Sendyk M.,  2000, \mn@doi [The Astrophysical Journal]
  {10.1086/309394}, 541, 351–366

\bibitem[\protect\citeauthoryear{{Gil}, {Melikidze}  \& {Geppert}}{{Gil}
  et~al.}{2003}]{Gil_2003}
{Gil} J.,  {Melikidze} G.~I.,   {Geppert} U.,  2003, \mn@doi [\aap]
  {10.1051/0004-6361:20030854}, \href
  {https://ui.adsabs.harvard.edu/abs/2003A&A...407..315G} {407, 315}

\bibitem[\protect\citeauthoryear{Gil, Melikidze  \& Zhang}{Gil
  et~al.}{2006}]{Gil_2006}
Gil J.,  Melikidze G.,   Zhang B.,  2006, \mn@doi [The Astrophysical Journal]
  {10.1086/506982}, 650, 1048

\bibitem[\protect\citeauthoryear{{Hurley-Walker} et~al.,}{{Hurley-Walker}
  et~al.}{2024}]{ULP_2.9h}
{Hurley-Walker} N.,  et~al., 2024, \mn@doi [\apjl] {10.3847/2041-8213/ad890e},
  \href {https://ui.adsabs.harvard.edu/abs/2024ApJ...976L..21H} {976, L21}

\bibitem[\protect\citeauthoryear{{Johnston} \& {Karastergiou}}{{Johnston} \&
  {Karastergiou}}{2017}]{ppdot_evolution_Johnston}
{Johnston} S.,  {Karastergiou} A.,  2017, \mn@doi [\mnras]
  {10.1093/mnras/stx377}, \href
  {https://ui.adsabs.harvard.edu/abs/2017MNRAS.467.3493J} {467, 3493}

\bibitem[\protect\citeauthoryear{{Johnston} \& {Karastergiou}}{{Johnston} \&
  {Karastergiou}}{2019}]{duty_vs_period_johnston}
{Johnston} S.,  {Karastergiou} A.,  2019, \mn@doi [\mnras]
  {10.1093/mnras/stz400}, \href
  {https://ui.adsabs.harvard.edu/abs/2019MNRAS.485..640J} {485, 640}

\bibitem[\protect\citeauthoryear{{Johnston} \& {Kerr}}{{Johnston} \&
  {Kerr}}{2018}]{EPN_1379}
{Johnston} S.,  {Kerr} M.,  2018, \mn@doi [\mnras] {10.1093/mnras/stx3095},
  \href {https://ui.adsabs.harvard.edu/abs/2018MNRAS.474.4629J} {474, 4629}

\bibitem[\protect\citeauthoryear{{Johnston}, {Kramer}, {Karastergiou}, {Keith},
  {Oswald}, {Parthasarathy}  \& {Weltevrede}}{{Johnston}
  et~al.}{2023}]{TPA_RVM1}
{Johnston} S.,  {Kramer} M.,  {Karastergiou} A.,  {Keith} M.~J.,  {Oswald}
  L.~S.,  {Parthasarathy} A.,   {Weltevrede} P.,  2023, \mn@doi [\mnras]
  {10.1093/mnras/stac3636}, \href
  {https://ui.adsabs.harvard.edu/abs/2023MNRAS.520.4801J} {520, 4801}

\bibitem[\protect\citeauthoryear{{Johnston}, {Mitra}, {Keith}, {Oswald}  \&
  {Karastergiou}}{{Johnston} et~al.}{2024}]{TPA_RVM2}
{Johnston} S.,  {Mitra} D.,  {Keith} M.~J.,  {Oswald} L.~S.,   {Karastergiou}
  A.,  2024, \mn@doi [\mnras] {10.1093/mnras/stae1175}, \href
  {https://ui.adsabs.harvard.edu/abs/2024MNRAS.530.4839J} {530, 4839}

\bibitem[\protect\citeauthoryear{{Karastergiou} \& {Johnston}}{{Karastergiou}
  \& {Johnston}}{2007}]{profile_complexity_ArisK}
{Karastergiou} A.,  {Johnston} S.,  2007, \mn@doi [\mnras]
  {10.1111/j.1365-2966.2007.12237.x}, \href
  {https://ui.adsabs.harvard.edu/abs/2007MNRAS.380.1678K} {380, 1678}

\bibitem[\protect\citeauthoryear{Kramer, Johnston  \& Van~Straten}{Kramer
  et~al.}{2002}]{Kramer_2002}
Kramer M.,  Johnston S.,   Van~Straten W.,  2002, \mn@doi [Monthly Notices of
  the Royal Astronomical Society] {10.1046/j.1365-8711.2002.05478.x}, 334,
  523–532

\bibitem[\protect\citeauthoryear{{Kramer}, {Liu}, {Desvignes}, {Karuppusamy}
  \& {Stappers}}{{Kramer} et~al.}{2024}]{micro_kramer}
{Kramer} M.,  {Liu} K.,  {Desvignes} G.,  {Karuppusamy} R.,   {Stappers} B.~W.,
   2024, \mn@doi [Nature Astronomy] {10.1038/s41550-023-02125-3}, \href
  {https://ui.adsabs.harvard.edu/abs/2024NatAs...8..230K} {8, 230}

\bibitem[\protect\citeauthoryear{{Lee} et~al.,}{{Lee} et~al.}{2025}]{NS_ULP}
{Lee} Y.~W.~J.,  et~al., 2025, \mn@doi [arXiv e-prints]
  {10.48550/arXiv.2501.09133}, \href
  {https://ui.adsabs.harvard.edu/abs/2025arXiv250109133L} {p. arXiv:2501.09133}

\bibitem[\protect\citeauthoryear{{Li} et~al.,}{{Li} et~al.}{2024}]{ULP_44m}
{Li} D.,  et~al., 2024, \mn@doi [arXiv e-prints] {10.48550/arXiv.2411.15739},
  \href {https://ui.adsabs.harvard.edu/abs/2024arXiv241115739L} {p.
  arXiv:2411.15739}

\bibitem[\protect\citeauthoryear{{Lyne}, {Manchester}  \& {Taylor}}{{Lyne}
  et~al.}{1985}]{luminosity_lyne}
{Lyne} A.~G.,  {Manchester} R.~N.,   {Taylor} J.~H.,  1985, \mn@doi [\mnras]
  {10.1093/mnras/213.3.613}, \href
  {https://ui.adsabs.harvard.edu/abs/1985MNRAS.213..613L} {213, 613}

\bibitem[\protect\citeauthoryear{{Maciesiak} \& {Gil}}{{Maciesiak} \&
  {Gil}}{2011}]{LBL_interpulse}
{Maciesiak} K.,  {Gil} J.,  2011, \mn@doi [\mnras]
  {10.1111/j.1365-2966.2011.19359.x}, \href
  {https://ui.adsabs.harvard.edu/abs/2011MNRAS.417.1444M} {417, 1444}

\bibitem[\protect\citeauthoryear{Maciesiak, Gil  \& Melikidze}{Maciesiak
  et~al.}{2012}]{Maciesiak_2012}
Maciesiak K.,  Gil J.,   Melikidze G.,  2012, \mn@doi [Monthly Notices of the
  Royal Astronomical Society] {10.1111/j.1365-2966.2012.21246.x}, 424, 1762

\bibitem[\protect\citeauthoryear{{Manchester} \& {Taylor}}{{Manchester} \&
  {Taylor}}{1977}]{RFM_book}
{Manchester} R.~N.,  {Taylor} J.~H.,  1977, {Pulsars}

\bibitem[\protect\citeauthoryear{Mitra}{Mitra}{2017}]{Mitra_2017}
Mitra D.,  2017, \mn@doi [Journal of Astrophysics and Astronomy]
  {10.1007/s12036-017-9457-6}, 38

\bibitem[\protect\citeauthoryear{{Mitra}, {Basu}, {Melikidze}  \&
  {Arjunwadkar}}{{Mitra} et~al.}{2020}]{single_park}
{Mitra} D.,  {Basu} R.,  {Melikidze} G.~I.,   {Arjunwadkar} M.,  2020, \mn@doi
  [\mnras] {10.1093/mnras/stz3620}, \href
  {https://ui.adsabs.harvard.edu/abs/2020MNRAS.492.2468M} {492, 2468}

\bibitem[\protect\citeauthoryear{{Morello} et~al.,}{{Morello}
  et~al.}{2020a}]{morello_12s}
{Morello} V.,  et~al., 2020a, \mn@doi [\mnras] {10.1093/mnras/staa321}, \href
  {https://ui.adsabs.harvard.edu/abs/2020MNRAS.493.1165M} {493, 1165}

\bibitem[\protect\citeauthoryear{{Morello}, {Barr}, {Stappers}, {Keane}  \&
  {Lyne}}{{Morello} et~al.}{2020b}]{ffa_morello}
{Morello} V.,  {Barr} E.~D.,  {Stappers} B.~W.,  {Keane} E.~F.,   {Lyne} A.~G.,
   2020b, \mn@doi [\mnras] {10.1093/mnras/staa2291}, \href
  {https://ui.adsabs.harvard.edu/abs/2020MNRAS.497.4654M} {497, 4654}

\bibitem[\protect\citeauthoryear{{Philippov}, {Timokhin}  \&
  {Spitkovsky}}{{Philippov} et~al.}{2020}]{PIC_2020}
{Philippov} A.,  {Timokhin} A.,   {Spitkovsky} A.,  2020, \mn@doi [\prl]
  {10.1103/PhysRevLett.124.245101}, \href
  {https://ui.adsabs.harvard.edu/abs/2020PhRvL.124x5101P} {124, 245101}

\bibitem[\protect\citeauthoryear{{Posselt} et~al.,}{{Posselt}
  et~al.}{2021}]{TPA_width}
{Posselt} B.,  et~al., 2021, \mn@doi [\mnras] {10.1093/mnras/stab2775}, \href
  {https://ui.adsabs.harvard.edu/abs/2021MNRAS.508.4249P} {508, 4249}

\bibitem[\protect\citeauthoryear{{Posselt} et~al.,}{{Posselt}
  et~al.}{2023}]{TPA_datarelease}
{Posselt} B.,  et~al., 2023, \mn@doi [\mnras] {10.1093/mnras/stac3383}, \href
  {https://ui.adsabs.harvard.edu/abs/2023MNRAS.520.4582P} {520, 4582}

\bibitem[\protect\citeauthoryear{{Radhakrishnan} \& {Cooke}}{{Radhakrishnan} \&
  {Cooke}}{1969}]{RVM}
{Radhakrishnan} V.,  {Cooke} D.~J.,  1969, \aplett, \href
  {https://ui.adsabs.harvard.edu/abs/1969ApL.....3..225R} {3, 225}

\bibitem[\protect\citeauthoryear{{Reddy} et~al.,}{{Reddy}
  et~al.}{2017}]{GMRT_GWB}
{Reddy} S.~H.,  et~al., 2017, \mn@doi [Journal of Astronomical Instrumentation]
  {10.1142/S2251171716410117}, \href
  {https://ui.adsabs.harvard.edu/abs/2017JAI.....641011R} {6, 1641011}

\bibitem[\protect\citeauthoryear{{Roy}, {Chengalur}  \& {Pen}}{{Roy}
  et~al.}{2018}]{post_corr}
{Roy} J.,  {Chengalur} J.~N.,   {Pen} U.-L.,  2018, \mn@doi [\apj]
  {10.3847/1538-4357/aad815}, \href
  {https://ui.adsabs.harvard.edu/abs/2018ApJ...864..160R} {864, 160}

\bibitem[\protect\citeauthoryear{{Ruderman} \& {Sutherland}}{{Ruderman} \&
  {Sutherland}}{1975}]{RS75}
{Ruderman} M.~A.,  {Sutherland} P.~G.,  1975, \mn@doi [\apj] {10.1086/153393},
  \href {https://ui.adsabs.harvard.edu/abs/1975ApJ...196...51R} {196, 51}

\bibitem[\protect\citeauthoryear{{Sautron}, {P{\'e}tri}, {Mitra}  \&
  {Dirson}}{{Sautron} et~al.}{2024}]{population2}
{Sautron} M.,  {P{\'e}tri} J.,  {Mitra} D.,   {Dirson} L.,  2024, \mn@doi
  [\aap] {10.1051/0004-6361/202451097}, \href
  {https://ui.adsabs.harvard.edu/abs/2024A&A...691A.349S} {691, A349}

\bibitem[\protect\citeauthoryear{{Singh}, {Roy}, {Panda}, {Bhattacharyya},
  {Morello}, {Stappers}, {Ray}  \& {McLaughlin}}{{Singh}
  et~al.}{2022}]{GHRSSIII}
{Singh} S.,  {Roy} J.,  {Panda} U.,  {Bhattacharyya} B.,  {Morello} V.,
  {Stappers} B.~W.,  {Ray} P.~S.,   {McLaughlin} M.~A.,  2022, \mn@doi [\apj]
  {10.3847/1538-4357/ac7b91}, \href
  {https://ui.adsabs.harvard.edu/abs/2022ApJ...934..138S} {934, 138}

\bibitem[\protect\citeauthoryear{{Singh}, {Gupta}  \& {De}}{{Singh}
  et~al.}{2024}]{intrinsic_micro}
{Singh} S.,  {Gupta} Y.,   {De} K.,  2024, \mn@doi [\mnras]
  {10.1093/mnras/stad3334}, \href
  {https://ui.adsabs.harvard.edu/abs/2024MNRAS.527.2612S} {527, 2612}

\bibitem[\protect\citeauthoryear{{Song} et~al.,}{{Song}
  et~al.}{2023}]{TPA_drifting}
{Song} X.,  et~al., 2023, \mn@doi [\mnras] {10.1093/mnras/stad135}, \href
  {https://ui.adsabs.harvard.edu/abs/2023MNRAS.520.4562S} {520, 4562}

\bibitem[\protect\citeauthoryear{{Spiewak} et~al.,}{{Spiewak}
  et~al.}{2020}]{superb_v}
{Spiewak} R.,  et~al., 2020, \mn@doi [\mnras] {10.1093/mnras/staa1869}, \href
  {https://ui.adsabs.harvard.edu/abs/2020MNRAS.496.4836S} {496, 4836}

\bibitem[\protect\citeauthoryear{Thorsett}{Thorsett}{1992}]{Thorsett_1992}
Thorsett S.~E.,  1992, \mn@doi [International Astronomical Union Colloquium]
  {10.1017/S0002731600154952}, 128, 143–146

\bibitem[\protect\citeauthoryear{Timokhin}{Timokhin}{2010}]{PIC_2010}
Timokhin A.~N.,  2010, \mn@doi [Monthly Notices of the Royal Astronomical
  Society] {10.1111/j.1365-2966.2010.17286.x}, 408, 2092

\bibitem[\protect\citeauthoryear{{Vivekanand} \& {Radhakrishnan}}{{Vivekanand}
  \& {Radhakrishnan}}{1980}]{Profilecomplexity_vivekananda}
{Vivekanand} M.,  {Radhakrishnan} V.,  1980, \mn@doi [Journal of Astrophysics
  and Astronomy] {10.1007/BF02714232}, \href
  {https://ui.adsabs.harvard.edu/abs/1980JApA....1..119V} {1, 119}

\bibitem[\protect\citeauthoryear{{Wahl}, {Rankin}, {Venkataraman}  \&
  {Olszanski}}{{Wahl} et~al.}{2023}]{J1503_325MHz}
{Wahl} H.,  {Rankin} J.,  {Venkataraman} A.,   {Olszanski} T.,  2023, \mn@doi
  [\mnras] {10.1093/mnras/stac3613}, \href
  {https://ui.adsabs.harvard.edu/abs/2023MNRAS.520..314W} {520, 314}

\bibitem[\protect\citeauthoryear{Young, Manchester  \& Johnston}{Young
  et~al.}{1999}]{young_1999}
Young M.,  Manchester R.,   Johnston S.,  1999, \mn@doi [Nature]
  {10.1038/23650}, 400, 848

\bibitem[\protect\citeauthoryear{{Young}, {Chan}, {Burman}  \& {Blair}}{{Young}
  et~al.}{2010}]{beam_evolution_young}
{Young} M.~D.~T.,  {Chan} L.~S.,  {Burman} R.~R.,   {Blair} D.~G.,  2010,
  \mn@doi [\mnras] {10.1111/j.1365-2966.2009.15972.x}, \href
  {https://ui.adsabs.harvard.edu/abs/2010MNRAS.402.1317Y} {402, 1317}

\bibitem[\protect\citeauthoryear{{van Heerden}, {Karastergiou}  \&
  {Roberts}}{{van Heerden} et~al.}{2017}]{search_framework}
{van Heerden} E.,  {Karastergiou} A.,   {Roberts} S.~J.,  2017, \mn@doi
  [\mnras] {10.1093/mnras/stw3068}, \href
  {https://ui.adsabs.harvard.edu/abs/2017MNRAS.467.1661V} {467, 1661}

\makeatother
\end{thebibliography}





\bsp	
\label{lastpage}
\end{document}